\documentclass[a4paper,preprint,12pt]{elsarticle}
\usepackage{amsmath,amsthm,amssymb}
\usepackage{graphicx}
\usepackage[left=1.0in, right=0.75in, top=1.0in, bottom=1.0in]{geometry}
\usepackage{lineno}

\journal{none}

\begin{document}
\linenumbers

\begin{frontmatter}

\title{Distorted wurtzite unit cells: Determination of lattice parameters of non-polar $a$-plane AlGaN and estimation of solid phase Al content.}

 \author[label1]{Masihhur R. Laskar}
 \author[label2]{Tapas Ganguli}
 \author[label1]{A. A. Rahman}
 \author[label1]{Amlan Mukherjee}
 \author[label1]{M. R. Gokhale}
 \author[label1]{Arnab Bhattacharya}

\address[label1]{Department of Condensed Matter Physics and Materials Science, Tata Institute of Fundamental Research, Homi Bhabha Road,
Mumbai 400005, India.}
\address[label2]{Raja Ramanna Center for Advanced Technology, Indore 425013, India.}

\begin{abstract}
Unlike $c$-plane nitrides, ``non-polar" nitrides grown in e.g. the $a$-plane or $m$-plane orientation encounter anisotropic in-plane strain due to the anisotropy in the lattice and thermal mismatch with the substrate or buffer layer. Such anisotropic strain results in a distortion of the wurtzite unit cell and creates difficulty in accurate determination of lattice parameters and solid phase group-III content ($x_{solid}$) in  ternary alloys. In this paper we show that the lattice distortion is orthorhombic, and outline a relatively simple procedure for measurement of lattice parameters of non-polar group III-nitrides epilayers from high resolution x-ray diffraction measurements. We derive an approximate expression for $x_{solid}$ taking into account the anisotropic strain. We illustrate this using data for $a$-plane AlGaN, where we measure the lattice parameters and estimate the solid phase Al content, and also show that this method is applicable for $m$-plane structures as well.
\end{abstract}

\end{frontmatter}

\section{Introduction}
Group III-nitrides semiconductors have potential applications in optoelectronics and microelectronics devices. Nitride semiconductors epilayers grown along the $(0001)$ $c$-axis of the wurtzite crystal structure suffer from strong undesirable spontaneous and piezoelectric polarization fields, which give rise to internal electrical fields \cite{1} and impair device performance. In quantum wells these fields spatially separate the electrons and holes reducing the overlap of their wave functions, and causing a reduction of the recombination efficiency and red-shift of the emission peak in light-emitting devices \cite{2}. A solution to avoid the deleterious polarization-induced electric field effects is to use group-III nitride layers in crystal orientations which have no polarization field in the growth direction, and hence across the device active region \cite{paskova}. Therefore, there is extensive ongoing research towards the growth of ``non-polar" $(11\bar{2}0)$ $a$-plane and $(1\bar{1}00)$ $m$-plane ($\perp c$-axis) group III-nitrides. These non-polar $a$- or $m$-plane epilayers are generally grown on $r$- or $m$-plane sapphire substrates respectively. The lattice mismatch and thermal expansion coefficients of these nitride epilayers with respect to the substrate are different along $\parallel c$ and  $\perp c$-directions. This gives rise to an \emph{anisotropic} in-plane strain which distorts the basal-plane of the hexagonal unit cell. Similar anisotropic differences in lattice mismatch and expansion coefficients also exist within the different members of the III-Nitride family, hence there is an anisotropic in-plane strain even when growing for example an $a$-plane AlGaN epilayer on a $a$-plane GaN buffer layer. The distortion of the basal plane has also been observed for similar reason in case of $c$-plane GaN grown on $a$-plane sapphire substrate \cite{darak2002}.

In $c$-plane oriented films, epilayers are under uniform in-plane strain, which deforms the unit cell but preserves the hexagonal symmetry of the basal plane. In that case the determination of lattice parameter and estimation of group III molar fraction in the ternary alloy (for example Al content of AlGaN films) is relatively straightforward \cite{Angerer,Moram}. However, for the non-polar orientations, the anisotropic in-plane strain results a distortion of the wurtzite unit cell which breaks the hexagonal symmetry.  Further, the orientation of the $a$-plane nitride epilayer on $r$-plane sapphire substrate results in a reduced number of available lattice points in the reciprocal space within the limiting-sphere that are accessible for diffraction measurements. This makes determination of lattice parameters for non-polar nitrides and consequently the estimation of group III composition of non-polar nitrides very difficult \cite{darak,roder}. A few different procedures for lattice parameter determination of such structures have been discussed in the literature. Darakchieva et al. \cite{darak} detail a procedure that requires measuring several symmetric and skew-symmetric planes at multiple azimuth positions in an edge-symmetric geometry. Another approach by Roder et al.\cite{roder} uses measurements of interplanar spacings derived from a combination of 9 symmetric, asymmetric, and skew-symmetric reflections, which are weighted by their corresponding fwhm values and used in a fitting routine to match to an orthorhombic structure via an error minimization routine. In this paper we suggest a slightly different procedure for measuring the lattice parameters for such distorted systems using high resolution x-ray diffraction (HRXRD). We first show that the anisotropic strain results in an orthorhombic distortion of the unit cell, and derive a general expression for the interplanar distance $d_{hkl}$ in such structures. Using the interplanar distances determined from multiple symmetric and skew-symmetric reflections, we obtain the lattice parameters via a standard least-square error minimization routine that is easily implemented in standard mathematical software packages using a matrix formulation. We also derive an approximate expression for the ternary alloy composition of Al$_{x}$Ga$_{1-x}$N epilayers, where the Al content $x_{solid}$ is obtained solving the stress-strain tensor taking into account the anisotropic strain. The procedure is illustrated using measurements on $a$-plane AlGaN epilayers grown on AlN buffer layers on r-plane sapphire substrates. We also show that the procedure is applicable for $m$-plane nitrides as well.

\section{Experiment}
The AlGaN  epilayers were grown via metal organic vapour phase epitaxy (MOVPE) in a $3\times2''$ closed-coupled showerhead reactor. Trimethylgallium (TMGa), trimethylaluminium (TMAl), and ammonia (NH$_3$) are used as precursors and Pd-diffused hydrogen (H$_2$) as carrier gas. About $0.8 \mu m$ thick AlGaN epilayers were grown on AlN buffer layer. The details of the growth procedure can be found in Ref. \cite{laskarPSS}. The lattice parameters measurement were carried out by using a Philips X'PERT$^{TM}$ high resolution X-ray diffractometer with a symmetric Ge($220$) hybrid monochromator and an asymmetric triple-axis analyzer and PIXcel solid state detector array. The 2$\theta$ value of a set of planes $(11\bar{2}0)$, $(2\bar{1}\bar{1}0)$, $(\bar{1}2\bar{1}0)$, $(10\bar{1}0)$, $(21\bar{3}0)$, $(2\bar{1}\bar{1}0)$, $(11\bar{2}0)$, $(10\bar{1}1)$, $(10\bar{1}2)$, $(21\bar{3}1)$, $(21\bar{3}2)$ were measured to confirm the orthorhombic distortion and to estimate the lattice parameters. Absorption measurements were done on backside polished samples using a Cary $5000$ spectrophotometer to estimate the band gap of the epilayers.

\section{Measurement of lattice parameters}

Fig 1(a) shows a schematic diagram of the relative unit cell orientation of an $a$-plane III-nitride epilayer on $r$-plane sapphire. The in-plane epitaxial relationships between the group III-nitride layer and sapphire are $[0001]_{nitride}$ $\parallel$ $[\bar{1}101]_{sapphire}$ and $[1\bar{1}00]_{nitride}$ $\parallel$ $[11\bar{2}0]_{sapphire}$. The thermal expansion co-efficients in the respective directions and the lattice mismatch are shown in the table 1. It is evident that the lattice and thermal mismatch along $[0001]$ and $[1\bar{1}00]$ are different which gives rise to the anisotropic strain in the overlayer and thus distorts the basal plane of the unit cell as shown schematically in Fig 1(b-c) (solid line). Further, the thermal mismatch along $c$ ($|\triangle \alpha_{[0001]}|$) and along $m$ ($|\triangle \alpha_{[1\bar{1}00]}|$) are larger for GaN/AlN compared to the GaN/sapphire and AlN/sapphire cases.

\begin{figure}[!h]
\centerline{\includegraphics*[width=14cm]{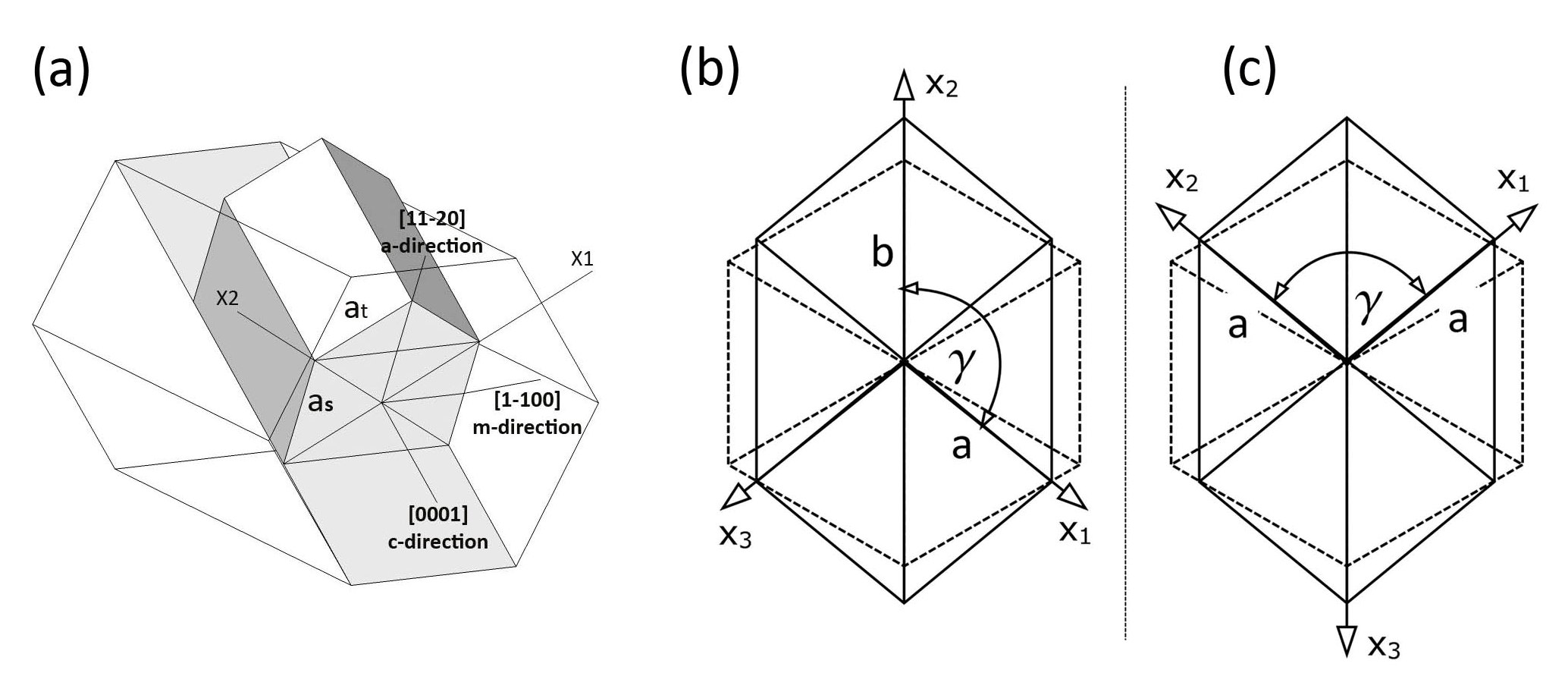}}
\caption{\label{fig1} (a) Schematic diagram showing the orientation of wurtzite unit cell of $(11\bar{2}0)$ $a$-plane oriented nitride epilayer on $(1\bar{1}01)$ $r$-plane sapphire substrate. Anisotropic in-plane strain results orthorhombic distortion and requires at least three lattice parameters $a_s$ (sidewall), $a_t$ (top) and $c$ for complete description of the unit cell. (b-c) The dotted lines and solid lines show the basal plane of a perfect and distorted hexagonal unit cell respectively. (b) A choice of coordinate axis that involves four independent lattice variables $a$, $b$, $c$ and $\gamma$,  whereas in (c) symmetry allows the reduction of one parameter ($a$=$b$) and gives a simpler expression for $d_{hkl}$ and also helps to visualize the lattice points in the reciprocal space.}
\end{figure}

\begin{table}[ht]
\caption{Thermal expansion and lattice mismatch}
\centering
\begin{tabular}{|c|c|c|c|}
\hline\hline
\emph{Thermal expansion} & \emph{GaN}          & \emph{AlN}          & \emph{Sapphire}         \\
\emph{coefficient}      &    $(10^{6} K^{-1}$)       &    $(10^{6} K^{-1}$)       &    $(10^{6} K^{-1}$)            \\
\hline\hline
$\alpha_{[0001]}$          & $3.17$        & $5.27$        & -                         \\ \hline
$\alpha_{[1\bar{1}00]}$    & $5.59$        & $4.15$        & -                          \\ \hline
$\alpha_{[\bar{1}101]}$    & -                         & -                         & $4.7$           \\ \hline
$\alpha_{[11\bar{2}0]}$    & -                         & -                         & $4.5$            \\ \hline
\end{tabular}
\begin{tabular}{|c|c|c|c|c|c|c|}
\hline
\emph{Lattice} & \multicolumn{2}{|c|}{\emph{\% w.r.t. GaN}} & \multicolumn{2}{|c|}{\emph{\% w.r.t. AlN}} & \multicolumn{2}{|c|}{\emph{\% w.r.t. Sapphire}} \\
\cline{2-3} \cline{4-5} \cline{6-7}
\emph{mismatch}   &  along $c$ & along $m$    &    along $c$ & along $m$        &   along $c$ & along $m$            \\ \hline\hline
\emph{GaN}        & 0          & 0            & 4.09         & 2.45             & 1.01        & 16.06                  \\ \hline
\emph{AlN}        & -3.92      & -2.39        & 0            & 0                & -2.87       & 13.29                  \\ \hline\hline
\end{tabular}
\end{table}

\subsection{Choice of coordinate axis}
For a perfect hexagonal unit cell the inter-planar distance $d_{hkl}$ (or $d_{hkil}$ where $i=-h-k$) between the $(hkl)$-planes is given by $1/d_{khl}^2=4/3.[(h^2+k^2+hk)/a^2]+l^2/c^2$. This expression cannot be used for the distorted structure shown in Fig.1(b or c) (solid line). To obtain a relatively simpler expression for new $d_{hkl}$, a proper choice of coordinate axis is helpful. In the first choice [Fig 1(b)], the inter-planar lattice distance involves four independent lattice variables $a$, $b$, $c$, and $\gamma$, so the expression of the $d_{hkl}$ will be complicated. But for the second choice [Fig. 1(c)], the symmetry in the distorted basal plane allows one to use $a$=$b$ and reduce one parameter thus simplifying the expression for the $d_{hkl}$. Further, such choice of coordinate axis helps to visualize the lattice points in the reciprocal space, as discussed later in section 3.3.

For the most general unit cell i.e. a triclinic structure, (Lattice sides $a\neq b\neq c$ and  angles $\alpha \neq \beta \neq \gamma$) the inter-planar lattice distance can be described by \cite{Cullity}

\begin{align*}
\frac{1}{d_{hkl}^{2}}=\frac{1}{V^2}(S_{11}h^2+S_{22}k^2+S_{33}l^2+2 S_{12}hk+S_{23}kl+S_{13}hl)\\
V=abc\sqrt{1-\cos^2\alpha-\cos^2\beta-\cos^2\gamma-2\cos\alpha \cos\beta \cos\gamma}\\
S_{11}=b^2c^2 \sin^2\alpha \\
S_{22}=a^2c^2\sin^2\beta  \\
S_{33}=a^2b^2\sin^2\gamma \\
S_{12}=abc^2(\cos\alpha \cos\beta-\cos\gamma) \\
S_{23}=a^2bc(\cos\beta \cos\gamma-\cos\alpha) \\
S_{13}=ab^2c(\cos\beta \cos\gamma-\cos\alpha) \\
\end{align*}

For a hexagonal lattice with orthorhombic distortion, we can put $\alpha=\beta=90^{o}$ and $a=b$, and thus obtain a simplified form

\begin{equation}
\frac{1}{d_{hkl}^{2}}=\frac{h^2+k^2-2hk\cos\gamma}{(a\sin\gamma)^2}+\frac{l^2}{c^2}
\end{equation}

The above expression for $d_{hkl}$ is similar to that for an undistorted hexagonal structure, with additional terms $\cos(\gamma)$ and $\sin(\gamma)$ which take into accounts the distortion of the basal plane.

\subsection{Accuracy in $d_{hkil}$ estimation}

Throughout the experiment, the $2\theta$ value for each plane was measured in the triple axis geometry and the interplanar distance $d_{hkil}$ was estimated from the Bragg condition given by $d_{hkil}=\lambda/2\sin\theta_{hkil}$. If $\bigtriangleup\theta$ is the error in the determination of peak position then the corresponding error in estimation of $d$ is given by $\bigtriangleup d=-d\cot\theta \bigtriangleup\theta$. The error $\bigtriangleup\theta$ can be minimized by careful optimization of the Eulerian cradle as discussed in \cite{Fewster}. To check the consistency in our measurements we repeated the measurement of the $2\theta$ value for the $(11\bar{2}0)$ reflection several times by taking out the sample and reloading it. It was found that the change in $2\theta$-value occurs in the third decimal place, which would lead to a corresponding change in the $d$ value, $\bigtriangleup d/d$, of order $10^{-5}$. All our measurement are based on the accuracy of $d_{hkl}$-values up to the 4th decimal place. Further, we have not incorporated a refractive index correction, and changes in $d_{hkl}$-values due to temperature fluctuations during the measurement (within $5^{o}$ C) which both result in changes of $\bigtriangleup d/d$, of order $10^{-5}$.


\subsection{Confirmation of orthorhombic distortion}

We outline below a quick way to confirm the nature of distortion of the unit cell to be orthorhombic. We choose the symmetric $(11\bar{2}0)$ plane and its equivalent planes $(2\bar{1}\bar{1}0)$ and $(\bar{1}2\bar{1}0)$ which are easily accessible in skew-symmetric geometry by adjusting the rotational axis $\phi$ and $\psi$ ($\phi$ is the angle of rotation about the normal to the sample mounting surface and $\psi$ is the tilt of the diffracting plane out of the diffractometer plane). From the symmetry of the wurtzite basal plane we expect that $d_{11\bar{2}0}$=$d_{2\bar{1}\bar{1}0}$=$d_{\bar{1}2\bar{1}0}$ for an undistorted hexagon as shown in Fig. 2a. Our measurements on a-plane GaN show that $d_{11\bar{2}0}$ $\geq$ $d_{2\bar{1}\bar{1}0}$=$d_{\bar{1}2\bar{1}0}$ as schematically shown in Fig. 2b(Data Shown in Table. 2). This implies that the unit cell gets compressed along $m$-direction and elongated along $a$-direction. Hence the angle between the axis vector $\mathbf{x_1}$ and $\mathbf{x_2}$ is less than $120^{o}$. For $a$-plane epilayers the direct measurement of $d_{1\bar{1}00}$ and $d_{\bar{1}100}$ is not possible because of geometry limitations of our diffractometer.

\begin{figure}[!h]
\centerline{\includegraphics*[width=12cm]{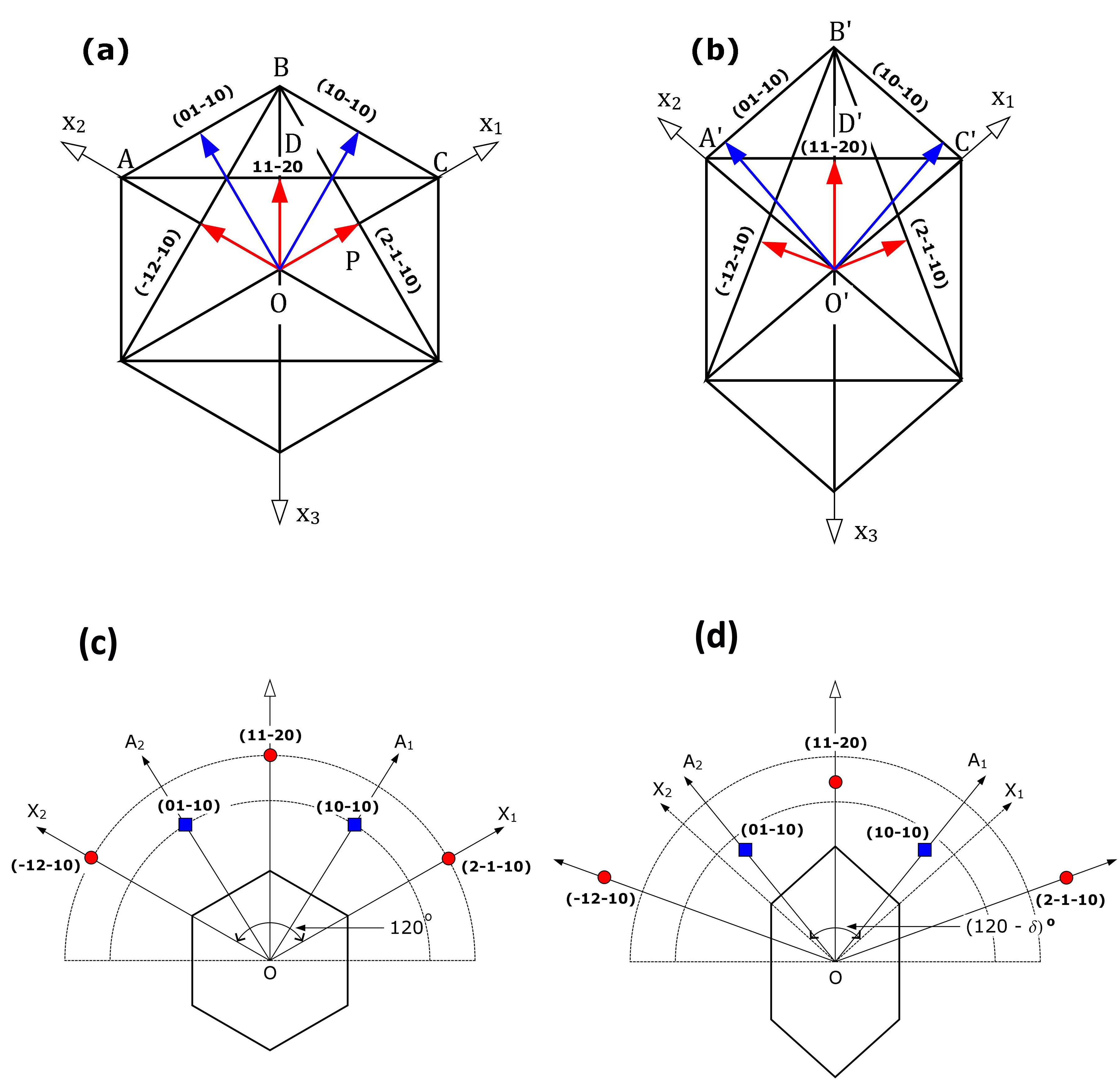}}
\caption{The diagram showing the basal plane of (a) perfect hexagonal and (b) orthorhombic distorted wurtzite unit cell. Different interplanar distance have been marked by arrows for both the structures, showing the inequalities. Reciprocal lattice points of (c) perfect hexagonal and (d) orthorhombic distorted structures.}
\end{figure}

Further, the two skew-symmetric planes $(10\bar{1}0)$, $(01\bar{1}0)$ give identical $2\theta$-values (i.e. $d_{10\bar{1}0} = d_{01\bar{1}0}$). Since $d_{10\bar{1}0}$ is the interplanar distance between BC and AO planes (Fig. 2), and  $d_{01\bar{1}0}$ is the same between AB and OC -planes, so we can write OA=OC or $a=b$. We next choose a set of asymmetric planes which involve the $c$ lattice parameter, like $(11\bar{2}2)$, $(11\bar{2}\bar{2})$; $(10\bar{1}1)$, $(10\bar{1}\bar{1})$, $(01\bar{1}1)$, $(01\bar{1}\bar{1})$; $(2\bar{1}\bar{1}2)$, $(\bar{1}2\bar{1}2)$; we found that each set of asymmetric planes gives identical $2\theta$-value (within measurements error: $\triangle\theta\pm0.001^{o}$). Further, it is also found that the peak positions of the $(10\bar{1}0)$ and $(01\bar{1}0)$ have the azimuthal relationship $\phi_{10\bar{1}0}=\pm 180^{o}+ \phi_{01\bar{1}0}$, similar relationship holds true for $(2\bar{1}\bar{1}0)$, $(\bar{1}2\bar{1}0)$ planes also. These measurements confirm that $\alpha = \beta = 90 ^{o}$. From these observations, we conclude that the hexagonal unit cell has orthorhombic distortion.

\begin{table}[ht]
\caption{Confirmation of orthorhombic distortion.}
\centering
\begin{tabular}{c c c c}
\hline\hline
Planes    & \multicolumn{2}{c}{Measured value (distorted)}   & Undistorted  \\
\cline{2-3}
(hkil)& 2$\theta_{hkil}$ (degree) & $d_{hkil}$ (\AA)& ($d'_{hkil}$) (\AA) \\ [1ex]
\hline \\
$(11\bar{2}0)$        & 57.6378 & 1.5979 & 1.5947  \\
$(2\bar{1}\bar{1}0)$  & 57.9779 & 1.5894 & 1.5947 \\
$(\bar{1}2\bar{1}0)$  & 57.9769 & 1.5894 & 1.5947 \\[1ex]
\hline
\end{tabular}
\label{table:nonlin}
\end{table}

\subsection{Least square method}
For an orthorhombic distorted hexagonal unit cell, we have derived an expression for the inter-planar distance $d_{hkl}$ (Eq.1). Since this expression has only three independent variables $a$, $c$, and $\gamma$, these can be evaluated in principle by measuring the $2\theta$-values of only three reflections of non-equivalent planes  [for example $(11\bar{2}0)$, $(10\bar{1}0)$ and $(10\bar{1}1)$] from which the lattice parameters can be estimated. But to improve the accuracy of measurement it is better to measure multiple reflections and minimize the error using a least square method.

Here we can write $\gamma=120^{o} - \delta$. If $|\delta|$ is small ($\leq 1^{o}$), we can write $\sin(\gamma)=\sin(120^{o}-\delta)=1/2(1-\sqrt{3}\delta)$ and $\cos(\gamma)=\cos(120^{o}-\delta)=\sqrt{3}/{2}(1-\delta/{\sqrt{3}})$. Substituting back to the eq(1), we can write an approximate expression:

\begin{eqnarray*}
\frac{1}{d_{hkl}^2}&=&\frac{h^2+k^2+hk(1+\sqrt{3}\delta)}{\frac{3}{4}a^2\left(1-\frac{\delta}{\sqrt{3}}\right)^2}+\frac{l^2}{c^2}\\
&=& \frac{4}{3}\frac{h^2+k^2+hk(1+\sqrt{3}\delta)}{a^2} \left(1-\frac{\delta}{\sqrt{3}}\right)^{-2}+\frac{l^2}{c^2}\\
&=& \frac{4}{3}\frac{h^2+k^2+hk(1+\sqrt{3}\delta)}{a^2} \left(1+\frac{2\delta}{\sqrt{3}}\right)+\frac{l^2}{c^2} \\
&=& \frac{4}{3}\frac{(h^2+k^2+hk)}{a^2} + \frac{4}{3\sqrt{3}}\frac{(2h^2+2k^2+5hk)}{a^2}\delta+ \frac{l^2}{c^2}
\end{eqnarray*}

Here we have neglected the higher order $\delta ^2$-term. By rearranging it, we can write the expression in the form of 3 linear variables:

\begin{equation}
\frac{1}{d_{hkl}^{2}}=\left[\frac{4}{3}(h^2+k^2+hk)\right].\frac{1}{a^2} + \left[\frac{4}{3\sqrt{3}}(2h^2+2k^2+5hk)\right].\frac{\delta}{a^2} + \left[l^2\right].\frac{1}{c^2}
\end{equation}

Assuming $x_1=1/a^2$, $x_2=\delta/a^2$ and $x_3=1/c^2$ we can express Eqn.(2) in a linear form with variables $x_1$, $x_2$ and $x_3$. Now for a set of $n$ reflecting planes say, $(h_1, k_1, l_1)$, $(h_2, k_2, l_2)$ ....$(h_n, k_n, l_n)$; we will get $n$-equations which can be expressed in a matrix form \textbf{Ap=D}, where \textbf{A} is a $(3\times n)$ matrix whose element $[A]_{qj}$ is the co-efficients of the variable $x_{j}$ $(j=1,2,3)$ of the $q$-th equation $(q=1,2,...n)$, \textbf{p}=($x_1$ $x_2$ $x_3$), and \textbf{D}=($1/d^2_{1}$  $1/d^2_{2}$ ... $1/d^2_{n}$)$^T$, where $d_{n}=d_{h_n k_n l_n}$ and T denotes transpose of the matrix. The required matrix, \textbf{p}, whose elements contain the lattice parameters can be obtained by solving the matrix equation \textbf{p=(A$^{T}$A)$^{-1}$ (A$^{T}$D)}. This is of a form that can be easily implemented in standard software packages like Mathematica and Matlab.

\subsection {Example}

For our $a$-plane GaN on AlN buffer layer, we have measured the 2$\theta$-value of a set of nine reflections as shown in Table 3 and then calculated the matrix \textbf{A} and \textbf{D}.

\vspace{1cm}

\begin{table}[ht]
\caption{2$\theta$-value for the set of planes for a-plane GaN}
\centering
\begin{tabular}{c c c}
\hline\hline
Planes &     $2\theta_{hkil}$ (degree)  &  d$_{hkil}$ (\AA) \\ [1ex]
\hline \\
$(11\bar{2}0)$              & 57.64     & 1.5980\\
$(10\bar{1}0)$              & 32.38     & 2.7628\\
$(21\bar{3}0)$              & 94.86     & 1.0459\\
$(2\bar{1}\bar{1}0)$        & 57.98     & 1.5894\\
$(11\bar{2}2)$              & 69.00     & 1.3600\\
$(10\bar{1}1)$              & 36.85     & 2.4373\\
$(10\bar{1}2)$              & 48.13     & 1.8889\\
$(21\bar{3}1)$              & 97.41     & 1.0253\\
$(21\bar{3}2)$              & 105.18    & 0.9698\\[1ex]
\hline
\end{tabular}
\label{table:1}
\end{table}

\vspace{1cm}

$\textbf{A} = \left( \begin{matrix}
4.00 & -6.93  & 0.00 \\
1.33 & -1.54  & 0.00 \\
9.33 & -15.39 & 0.00 \\
4.00 & -0.00  & 0.00 \\
4.00 & -6.93  & 4.00 \\
1.33 & -1.54  & 1.00 \\
1.33 & -1.54  & 4.00 \\
9.33 & -15.39 & 1.00 \\
9.33 & -15.39 & 4.00 \\
\end{matrix} \right)$ ; and  $\textbf{D} = \left( \begin{matrix} 0.39\\ 0.13\\0.91\\0.39\\0.54\\0.17\\0.28\\0.95\\1.06 \end{matrix} \right)$.

\vspace{1cm}
By solving the matrix equation for \textbf{p} using Matlab, we obtain the solution for the lattice parameters as $a=3.1788$ \AA, $c=5.1774$ \AA, and with $\delta=0.36^{o}$, with standard deviations for $ \sigma (a)=2.9\times 10^{-9}$, $ \sigma (c)=6.2\times 10^{-9}$ and $ \sigma (\delta)=4.1\times 10^{-8}$ respectively. We note that the value of $\delta$ obtained is small and satisfies the assumption of $\delta \leq 1^{0}$. Following the same procedure we have estimated the lattice parameters for the set of $a$-plane AlGaN samples, details are shown in Table 4.


\subsection{Procedure for $m$-plane structure}
In case of $m$-plane nitrides are grown on the $m$-plane sapphire substrates. Since the lattice and thermal mismatch along $c$ and $a$-axis are different, the distortion in the basal plane will be similar to the $a$-plane nitrides.
Figure 3(a) shows the orientation of unit cell of $m$-plane nitride on $m$-plane sapphire substrate.

\begin{figure}[!h]
\centerline{\includegraphics*[width=14cm]{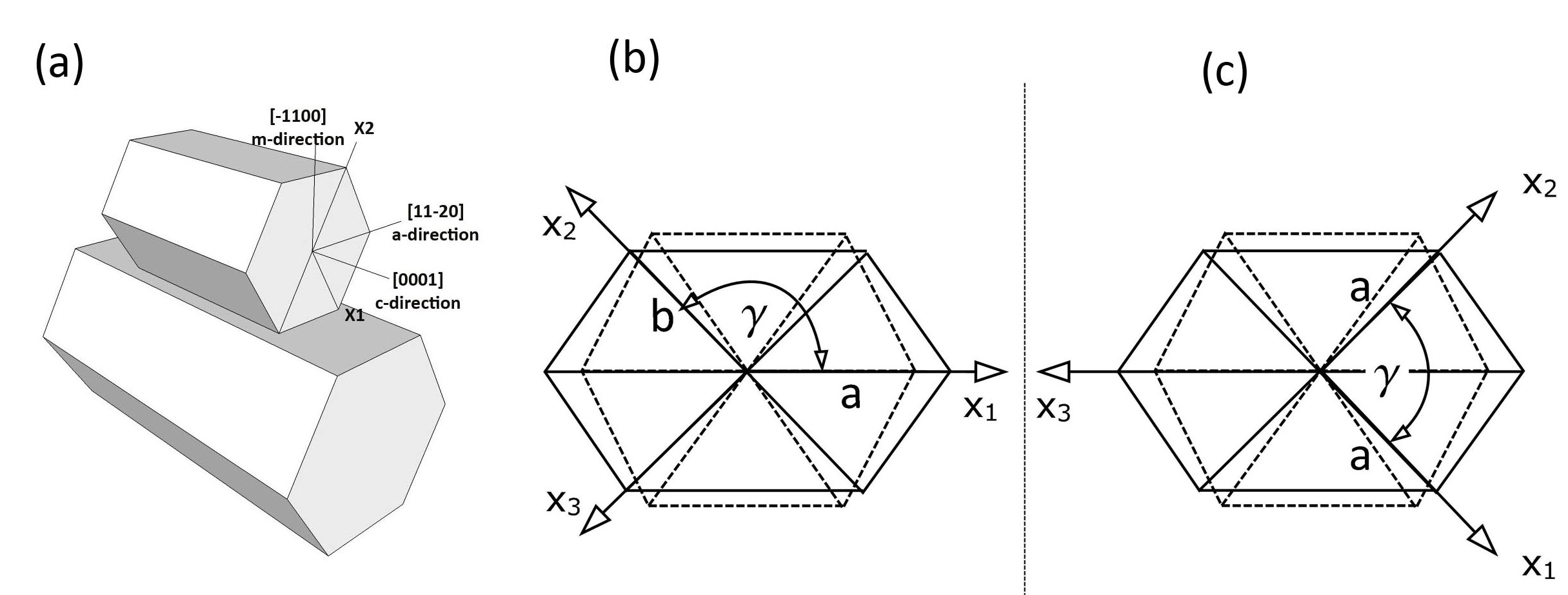}}
\caption{\label{fig1} (a) Schematic diagram showing the orientation of wurtzite unit cell of $(\bar{1}100)$ $m$-plane oriented nitride epilayer on $(\bar{1}100)$ $m$-plane sapphire substrate. Anisotropic in-plane strain results in orthorhombic distortion. The dotted lines and solid line shows the basal plane of a perfect and distorted hexagonal unit cell respectively. (b) A choice of coordinate axis that involves four independent lattice variables $a$, $b$, $c$ and $\gamma$,  whereas in (c) symmetry allows the reduction of one parameter ($a$=$b$) and gives a simpler expression for $d_{hkl}$.}
\end{figure}

As in the case of $a$-plane nitrides, an appropriate choice of coordinate axes can result in a simpler form of $d_{hkl}$. The first choice (Fig.3b), the inter-planar lattice distance involves four independent lattice variables $a$, $b$, $c$, and $\gamma$, whereas in the second choice (Fig. 3c), the symmetry in the distorted basal plane allows one to use $a$=$b$ and reduces one parameter resulting in a simpler expression for $d_{hkl}$ which is exactly identical to Eq(1). Also the distortion in the unit cell can be verified using a similar procedure by measuring if the inter-planar spacings follow $d_{\bar{1}100}$ $\leq$ $d_{\bar{1}010}$=$d_{01\bar{1}0}$.

This then allows the least square method as mentioned above for $a$-plane system to be used to estimate the lattice parameters for $m$-plane nitrides.

\section{Estimation of Al content}
For completely relaxed Al$_{x}$Ga$_{1-x}$N films the solid phase Al content ($x_{solid}$) can be estimated by  measuring either the $a$ or $c$ lattice parameter, subject to the validity of Vegard's law
\begin{align*}
a_{0}=x. a_{A}+ (1-x). a_{G} \tag{3a} \\
c_{0}=x. c_{A}+ (1-x). c_{G} \tag{3b}
\end{align*}
Here $(a_G,c_G)$ and $(a_A,c_A)$ are the lattice parameters of relaxed GaN and AlN respectively. Rewriting the above equations
\vspace{-1ex}
\begin{eqnarray*}
x=(a_{G}-a_{0})/(a_{G}-a_{A})  \\
x=(c_{G}-c_{0})/(c_{G}-c_{A})  \\
\end{eqnarray*}

\vspace{-3ex} Ideally, the $x$ value calculated by using either the $a$- or $c$-lattice parameters should be the same. But in practice for epitaxial AlGaN these values differ. As we have seen in \emph{Sec.3} the AlGaN epilayers have thermal and lattice mismatches with the buffer/substrate which cause deformation/distortion in the wurtzite unit cell. For accurate estimation of $x_{solid}$ using lattice parameters it is hence necessary to take into account the in-plane strain effect caused by buffer/substrate.

\subsection{stress-strain tensor matrix}
Considering the $X$-axis along $[11\bar{2}0]$, $Y$-axis along $[1\bar{1}00]$ and $Z$-axis along $[0001]$ direction, the strain-stress relation for hexagonal crystals with a $C_{v6}$ symmetry can be expressed as \cite{Gil}

\vspace{3ex}
$\begin{pmatrix}
\sigma_{xx}\\ \sigma_{yy} \\ \sigma_{zz} \\ \sigma_{xz} \\ \sigma_{yz} \\ \sigma_{xy} \\
\end{pmatrix} =
\begin{pmatrix}
C_{11} & C_{12} & C_{13} & 0      & 0      & 0      \\
C_{12} & C_{11} & C_{13} & 0      & 0      & 0      \\
C_{13} & C_{13} & C_{33} & 0      & 0      & 0      \\
0      & 0      & 0      & C_{44} & 0      & 0      \\
0      & 0      & 0      & 0      & C_{44} & 0      \\
0      & 0      & 0      & 0      & 0      & C_{66} \\
\end{pmatrix}
\begin{pmatrix}
\epsilon_{xx}\\ \epsilon_{yy} \\ \epsilon_{zz} \\ \epsilon_{xz} \\ \epsilon_{yz} \\ \epsilon_{xy} \\
\end{pmatrix}$
\vspace{3ex}

where $C_{ij}$ are the stiffness constants.

For $c$-plane nitrides system the crystal is free along $[0001]$-direction (or Z-axis), so $\sigma_{zz}=0$ and

\vspace{-3ex}
\begin{align*}
\epsilon_{zz}=-\frac{C_{13}}{C_{33}}(\epsilon_{xx}+\epsilon_{yy})  \tag{4a}
\end{align*}

For $a$-plane nitride epilayers the crystal is free along $[11\bar{2}0]$-direction (or $X$-axis), so $\sigma_{xx}=0$ and

\vspace{-3ex}
\begin{align*}
\epsilon_{zz}=-\frac{C_{11}}{C_{13}}\epsilon_{xx}-\frac{C_{12}}{C_{13}}\epsilon_{yy}  \tag{4b}
\end{align*}

For $m$-plane nitride epilayers the crystal is free along $[\bar{1}100]$-direction (or $Y$-axis), so $\sigma_{yy}=0$ and

\vspace{-3ex}
\begin{align*}
\epsilon_{zz}=-\frac{C_{12}}{C_{13}}\epsilon_{xx}-\frac{C_{11}}{C_{13}}\epsilon_{yy}  \tag{4c}
\end{align*}

\subsection{$c$-plane AlGaN: effect of deformation on determination of Al content}
We briefly review the well known case of $c$-plane oriented epilayers \cite{Angerer}, where the in-plane strain is isotropic which deforms the unit cell, but maintains the hexagonal symmetry. So the strain values are
\begin{align*}
\epsilon_{xx}=\epsilon_{yy}=(a-a_{0})/a_{0}, \hspace{3ex} \epsilon_{zz}=(c-c_{0})/c_{0}
\end{align*}

Substituting in eq(4a), we obtain

\begin{align*}
\frac{(a-a_{0})}{a_{0}}+ \gamma \frac{(c-c_{0})}{c_{0}}=0
\end{align*}

Where $\gamma=2C_{13}/C_{33}$. Substituting the expression for $a_{0}$ and $c_{0}$ and  for small strain values, $x_{solid}$ can be expressed in terms of lattice parameters as

\vspace{-3ex}
\begin{align*}
x=\frac{a(c_{G}-c)+\gamma (a_{G}-a)}{a(a_{G}-c_{A})+\gamma (c_{G}-c_{A})} \tag{5}
\end{align*}





\subsection{$a$-plane AlGaN: effect of deformation on determination of Al content}
As discussed in \emph{Sec.3} $a$-plane nitride epilayers have in-plane anisotropic strain which distorts the basal plane. Hence, it is necessary to incorporate the effect of anisotropic strain to obtain a correct expression for Al content.

\begin{figure}[!h]
\centerline{\includegraphics*[width=6cm]{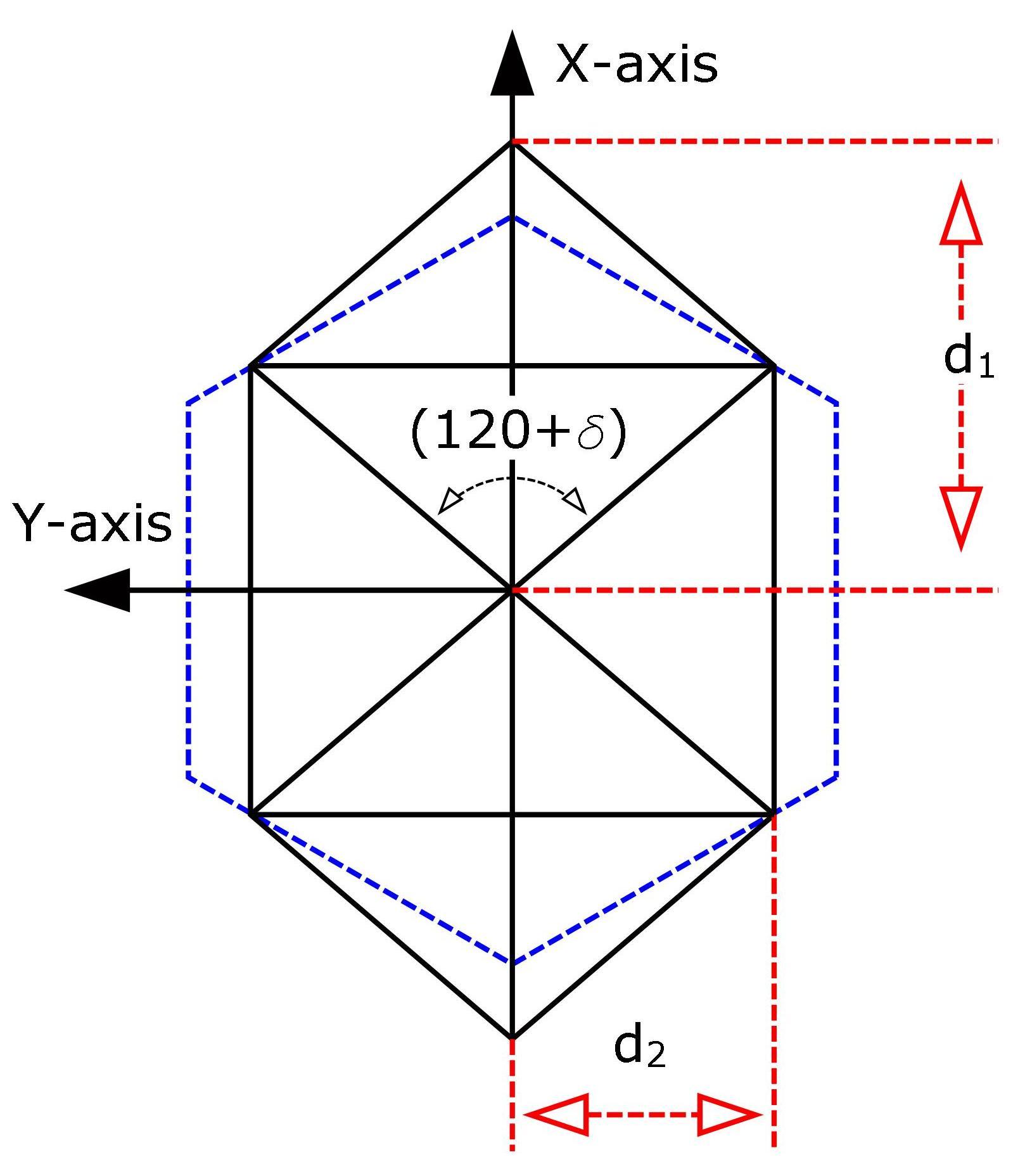}}
\caption{\label{fig1} The dotted curve showing the basal plane of perfect hexagon and the solid curve same for distorted unit cell. Here $1/2$ $d_1$ and $d_2$ are the interplanar distance of $(11\bar{2}0)$ and $(\bar{1}100)$ planes respectively. These values can be estimated for the expression of $d_{hkl}$ in eq.1}
\end{figure}

Here the strain values are

\begin{align*}
\epsilon_{xx}=(d_1-d_{10})/d_{10}\\
\epsilon_{yy}=(d_2-d_{20})/d_{20}\\
\epsilon_{zz}=(c-c_{0})/c_{0}\\
\end{align*}

where $d_{1}$=$2 d_{11\bar{2}0}$=$(2a\sin\gamma)/\sqrt{2(1-\cos\gamma)}$ and $d_{2}$=$d_{1\bar{1}00}$ = $(a\sin\gamma)/\sqrt{2(1+\cos\gamma)}$ are as indicated in the schematic diagram of the distorted basal plane in Fig. 4. The additional subscript `0' stands for relaxed AlGaN. Substituting these values into eqn. (4b) we obtain:

\begin{align*}
\frac{c-c_0}{c_0}+ \gamma_1 \frac{d_1-d_{10}}{d_{10}} + \gamma_2 \frac{d_2-d_{20}}{d_{20}}=0
\end{align*}

Assuming small strain and small distortion ($|\delta| \leq 1^{0}$), this we obtain

\begin{align*}
x=\frac{(c_{G}-c)d_{1}d_{2}+\gamma_{1}(d_{1G}-d_{1})cd_{2}+\gamma_{2}(d_{2G}-d_{2})cd_{1}}{(c_{G}-c_{A})d_{1}d_{2}+\gamma_{1}(d_{1G}-d_{1A})cd_{2}+\gamma_{2}(d_{2G}-d_{2A})cd_{1}} \tag{6} \end{align*}

where the additional subscripts `$G$' and `$A$' stand for GaN and AlN respectively, and $\gamma_1$=$C_{11}/C_{13}$ and $\gamma_2$=$C_{12}/C_{13}$. For $m$-plane AlGaN the expression of $x$ has a similar form except with $\gamma_1$=$C_{12}/C_{13}$ and $\gamma_2$=$C_{11}/C_{13}$.

\subsection{Example - solid phase Al composition for $a$-plane AlGaN}
Following the procedure described in the previous section, we have estimated the lattice parameters and solid phase Al composition for the series of $a$-plane AlGaN samples grown over the entire composition range. The details of the lattice parameters derived from HRXRD and the corresponding $x_{solid}$ calculated are shown in Table 4. This also shows the values of $x_{solid}$ derived independently from optical transmission measurements

\begin{table}[ht]
\caption{Comparison between $x$-values obtained from X-ray and transmission.}
\centering
\begin{tabular}{cccccccc}
\hline\hline
$x_{gas}$   & $a$    & $c$     & $\gamma$       & $d_1$   & $d_2$   & $x_{solid}$   & $x_{solid}$    \\
            &({\AA}) & ({\AA}) &  (degree)      & ({\AA}) & ({\AA}) &    (XRD)      & (Trans.) \\ [1ex]
\hline \\
0.0 & 3.1960     & 5.1785   & 119.64    & 3.2132    & 2.7628    & 0.00  & 0.00 \\
0.2 & 3.1783     & 5.1507   & 119.67    & 3.1940    & 2.7479    & 0.18  & 0.19 \\
0.4 & 3.1608     & 5.1050   & 119.68    & 3.1762    & 2.7328    & 0.38  & 0.39 \\
0.5 & 3.1535     & 5.0928   & 119.70    & 3.1654    & 2.7269    & 0.46  & 0.48 \\
0.7 & 3.1362     & 5.0532   & 119.78    & 3.1466    & 2.7130    & 0.66  & 0.66 \\
1.0 & 3.1116     & 4.9777   & 120.02    & 3.1107    & 2.6950    & 1.00  & 1.00 \\
\hline
\end{tabular}
\end{table}

The measurement shows that the value of $x_{solid}$ estimated from XRD and optical transmission agrees within $\pm 2\%$. The $x_{solid}$ is slightly lower than the $x_{gas}$ because of parasitic reaction between TMAl and NH$_3$. A detailed discussion of the variation of strain and distortion with Al-content, and its effect on the microstructure is discussed in references \cite{laskarPSS} and \cite{laskarJCG2}, and is not presented here as the emphasis of this work is to discuss the procedure rather than the results.

\section{Conclusion}
In conclusion, we have observed that the anisotropic in-plane strain results in an orthorhombic distortion in the wurtzite unit cell for non-polar $a$-plane nitrides. We have suggested a quick method for confirming such a distortion to be orthorhombic, and derived an expression for $d_{hkl}$ value for such distorted unit cells. We have also provide relatively simple procedure for estimation of accurate lattice parameters using multiple reflections and minimizing the error by a least square method. Since the orthorhombic distortion creates a difficulty for estimating group III content in ternary alloys, we have presented a technique which estimates the the correct Al content in $a$-plane AlGaN films taking into account  the effect of anisotropic strain. We have also shown that this method is equally applicable for $m$-plane nitrides as well. These procedure should be valuable to researchers working on a wide range of non-polar III-nitride epilayers.

\end{document}